# Spatiotemporal dynamics in demography-sensitive disease transmission: COVID-19 spread in NY as a case study

Joydeep Munshi, Indranil Roy and Ganesh Balasubramanian ⌐

*Department of Mechanical Engineering & Mechanics, Lehigh University, Bethlehem, PA 18015*

**Abstract**

The rapid transmission of the highly contagious novel coronavirus has been represented through several data-guided approaches across targeted geographies, in an attempt to understand when the pandemic will be under control and imposed lockdown measures can be relaxed. However, these epidemiological models predominantly based on training data employing number of cases and fatalities are limited in that they do not account for the spatiotemporal population dynamics that principally contributes to the disease spread. Here, a stochastic cellular automata enabled predictive model is presented that is able to accurate describe the effect of demography-dependent population dynamics on disease transmission. Using the spread of coronavirus in the state of New York as a case study, results from the computational framework remarkably agree with the actual count for infected cases and deaths as reported across organizations. The predictions suggest that an extended lockdown in some form, for up to 180 days, can significantly reduce the risk of a second wave of the outbreak. In addition, increased availability of medical testing is able to reduce the number of infected patients, even when less stringent social distancing guidelines and imposed. Equipping this stochastic approach with demographic factors such as age ratio, pre-existing health conditions, robustifies the model to predict the transmittivity of future outbreaks before they transform into an epidemic.

---

⌐ Corresponding author. Email: bganesh@lehigh.edu. Phone: +1-610-758-3784.
Address: Packard Laboratory 561, 19 Memorial Drive West, Bethlehem, PA 18015, USA.

The classical and most primitive mathematical model for epidemiology, *viz.*, susceptible-infectious-removed/recovered (SIR), introduced in 1927,[1] has undergone several modifications to simulate the transmission of complex diseases.[2] Nevertheless, for the model predictions to be accurate, it is still critical to identify the rates at which a select population is susceptible to transformations in its conditions from latent, to infected and, subsequently, to recovered. These parameters are predominantly local, depending on factors such as demographics, population mobility, and frequency of human interactions, *e.g.,* those experienced during the use of public transportation. While the SIR approaches are equipped to reproduce and predict the disease spread within a geographic region, these models do not consider (1) incidences of inter-personal contacts, (2) spatial distribution of the epidemic and (3) individual behaviors. Thus, only solving SIR inspired differential equations do not provide the desired accuracy for epidemic simulations, especially for a disease like novel coronavirus (COVID-19) that has an extremely high transmittivity.

The outbreak of the severe acute respiratory system coronavirus 2, *i.e.* SARS-CoV-2 that causes COVID-19 and first reported in Wuhan, China[3] in December 2019, has recently exploded into a pandemic. This global crisis has already threatened humanity, while deeply impacting the global economy[4]. SARS-CoV-2 is highly contagious and infamous for its rapid transference.[5] At the time of this writing, the reported cases have been increasing exponentially throughout North America and Europe. On the other hand, China recorded a peak in the number of positive COVID-19 cases around mid-February, but afterwards, notably, new cases have steadily decreased reaching almost zero by the end of March, as reported by the World Health Organization (WHO).[6] Per the 86th daily report of the WHO (April 15, 2020),[6] there are ~1,914,916 confirmed cases of infection and ~123,010 reported deaths worldwide, which massively under-predicts the actual count. Of these reported patients, ~578,268 are in the U.S., and over 23,000 have died,[6] with ~34% of the cases localized in the state of New York (NY). The strict social distancing and lockdown of non-life sustaining businesses that proved effective in controlling the spread of the disease in China,[7] inspired different states in the U.S. to impose similar strategies, with varying levels of rigor, from mid-March and extended the same till the end of April. However, since local-demographics specific disease spread model to quantitatively analyze the transmission dynamics are sparse, the effects of various mitigation and damage-limitation measures against the outbreak, are rather crude.[3,7,8,9,10,11,12]

To address the above challenges, we construct a spatiotemporally dynamic model based on the stochastic cellular automata (CA) approach,[13,14] that overcomes the shortcomings of the SIR and SEIRS models.[15,16,17] This CA-SLQIJR (susceptible-latent-quarantine-infectious-isolated-recovered) framework incorporates the potential consequences of population mobility and social distancing, to accurately describe the disease spread in NY, and provide quantitative predictions on future trends as the states in the U.S. prepare to gradually relax the curfew. The population density across NY dictates the probabilities of random local walks, while an added state, *viz*, quarantine (Q), which implies self-quarantine during the lockdown period, is introduced here that was nonexistent in previous approaches. The probabilities considered for the disease transmittivity are near-realistic by indirectly incorporating factual data on the population's age distribution, pre-existing health conditions, employment records, ethnicities and backgrounds, all down to the county level. The predictions demonstrate remarkable agreement with the actual data available from the Johns Hopkins University, as of 14[th] April 2020.,[18] As discussed below, the model suggests that extension of the current lockdown in some form (*e.g.,* expanded timelines or repeated implementations in multiple phases till a cure is identified) with gradual relaxation of the restrictions in communities where the susceptible population resides, will significantly reduce the risk of another potential outbreak.

***Imposed lockdown in NY*:** The transient data on the total infected cases and fatalities are normalized with respect to the number of infections and deaths per 100,000 people. Note that the first infection in every 100,000 people was reported on 2[nd] March 2020. We interrogate four different scenarios with the CA-SLQIJR model: (i) *no lockdown* ($\gamma = 0$), (ii) *partial lockdown* that commences from 17[th] March 2020 through the next 45 days with population mobility reduced to the local community and 'stay at home' being abided by only ~ 30% of the people ($\gamma = 0.3$), (iii) *strong lockdown,* with the same duration as (ii) but with 'stay at home' guideline followed by ~70% population ($\gamma = 0.7$), and (iv) *severe lockdown* for the same duration but with ≥95% population obeying the regulation ($\gamma = 0.95$).

The actual cumulative record on the number of affected people and the number of deaths per 100,000 aligns with the *strong lockdown* scenario (Figures 1A and 1B) as predicted by our model. The predictions for the noncumulative daily number of cases are provided in the Supplementary Information (Figure S1). The outbreak is noted to rise again between the 60[th] and 70[th] day for the *strong lockdown* and *severe lockdown* scenarios highlighting that even a small number of infected people can cause a spike in the number of infected cases within a few days after complete relaxation of the restricted interactions. On the other hand, the predicted counts for infection and deaths with the *partial lockdown* gradually approaches the *no lockdown* scenario, and the impact of the imposed restrictions is apparently limited since the number of infections do appreciably increase during the *partial lockdown* period. This is because the number of infections for partial lockdown are not reduced to a number which will have a significant impact on the spreading. Although during the imposed *strong* and *severe lockdowns*, the outbreak is restricted to considerably low counts, our model suggests that a single 45-day duration for the lockdown is insufficient to completely mitigate the disease transmission. A second wave of spread is reproduced by the model, resulting in a substantial rise in the fatalities. Pictorial representations of the spatial distribution of the disease in NY (Figures 1C-F) display the predicted hot spots at different time instants since the first reported infection. On day 15, in accord with the actual time series data (17th March 2020), lockdown state (Q) is imposed over the entire population mimicking a *strong lockdown* scenario. The model-identified hot spots concur with the zone of high COVID-19 cases,[19] while the projected predictions for the future spread of the disease, considering the lockdown is removed after 45 days, reveal several new high susceptibility zones for the outbreak because the global movements will resume after the relaxation of lockdown. When we include one infected person to a highly populated region such as New York city the model reflects the effect of population density driven spread at different regions throughout the state. The randomness of the cell movement is sampled over several simulations to reduce any noise due to the local random walk implemented by our CA model.

***Lockdown extension as a potentially effective strategy***: The extension of the lockdown in some form over several months (~ 5-6 months) has been proposed by certain U.S. governors and international leaders.[20,21] As reported in the literature,[22] sustained non-pharmaceutical interventions are necessary to limit the outbreak-driven mortality till vaccines are available (typical timeline ~18 months) for the susceptible crowd. Motivated by recent reports on a potential resurgence of COVID-19 in China, we examine (Figures 2A-B) the impact of the lockdown period (between 45 – 180 days), when ~70% of the NY population is observing self-quarantine. Note that the model assumes unavailability of vaccine in the simulation timeframe and that the immunity of the recovered patients will eventually diminish with time. Our model corroborates that an immature revoking of the lockdown, without exhaustively isolating the infectious (including infected and asymptomatic) population, is invariably followed by a pervasive reoccurrence of the disease. Assuming that an average recovered person becomes susceptible to COVID-19 after ~ 3-6 months, we find that a lockdown extending to ~ 180 days will significantly reduce the number of confirmed cases during that time (the distribution of the daily count is provided in Figure S2 (A-B) as supplementary information). The basic assumption of this model is that the normal pre-pandemic life will resume after the lockdown is lifted. The spike of daily cases just after lifting of premature lockdown is due to relaxation of imposed rules over a significant number of contagious people. Our CA-SLQIJR model being sensitive to parameters such as availability of the drugs and vaccines, immunity of recovered crowd etc., we believe that the model can offer accurate recommendations with dynamic demographic inputs.

***Travel restrictions inside the state:*** Mobility of the population over a geography (such as cities, states or countries) is one of the major driving causes for an epidemic spread. From Figures 1(C-F), we observe clustered regions throughout the NY state citing hots spots of the outbreak. Assuming 70% of the population are obeying the lockdown guidelines, we examine the effect of the movement of the remaining 30% population (*i.e.* if the 30% population is moving within the community a.k.a. *local mobility* or traveling long distances) on the disease transmission. Figures 2C and D present the cumulative number of infected cases and fatalities for four different situations where: (a) >95% of the mobile population are commuting within the local community (short distance travel) such as grocery stores and restraining from long distance travel, while at most 5% of them are moving across the state, (b) >75% are adhering to the imposed rules and plying within the local community, while ~ 25% are travelling long distances, (c) >50% are moving just within the local community, and (d) >50% of the mobile population are not adhering to the imposed rules (only <50% are restricting movement within local neighborhood). Note that it is assumed for all the situations, at any given time the 70% of the susceptible population, due to the imposed lockdown, is still practicing self-quarantine. The daily number of cases and deaths are presented in Figures S2 (C-D). It is evident from the results that restricted travel across cities within NY can noticeably mitigate the effect of the spread. Note that >95% of the movements when limited to the local communities can significantly reduce transmission evinced by the significantly lower cumulative count of infections (Figure 2C).

*Impact of available medical facility:* Medical facilities such as availability of testing kits, hospital beds and antiviral drugs are crucial to diagnose patients with COVID-19 symptoms. Although a surplus of these facilities is not feasible during an epidemic, at least sufficient availability of medical facilities can reduce the panic and instill confidence within a local population. In our model, the availability of these facilities is introduced by varying the probability of an infected person (state I) transitioning to isolation (state J) *i.e.*, those who test positive are receiving proper medical attention. As this probability of isolation increases, depicted in Figures 2(E-F), the total number of predicted infections and fatalities reduces significantly over the time. In presence of adequate medical facilities (*i.e., > 95%* of the infections can be tested) the cumulative count of infections and fatalities consistently decrease over a 200-day period (Figures 2(E-F)), superior to an imposed lockdown for a similar duration. Our predictions corroborate that enhanced access to medical facilities not only increase the confidence in the susceptible population but also can largely reduce the requirement of lockdown extensions, contributing to both human life and economy.

In summary, we present findings from a novel spatiotemporal model for disease transmission that enables indirect forecast on the effect of population dynamics on disease spread, with COVID-19 in NY as a case study, specific to the demographics. The cellular automata-based model enables understanding of the transmission dynamics and the predictions strongly concur with the actual patient counts as reported by various health organizations. Our results suggest that an extension of the current lockdown with gradual relaxation of the restrictions (over the next 180 days) can greatly reduce the risk of a second wave of the outbreak. An increased focus on supplementing medical facilities, including availability of test kits, will contribute to low number of infected patients and fatalities, even with less stringent social distancing guidelines. The analyses of the population mobility are equally instrumental towards implementing the mitigation strategies at early stages of an outbreak. We also present a scenario in the supplementary document (Figure S3), where a multi-stage lockdown can potentially limit the patient count and deaths due to this disease. By tuning the model parameters to account for dynamic demographic factors such as age ratios, pre-existing health conditions, employment status etc., transmittivity of future outbreaks can be predicted *a priori* before it evolves into an epidemic.

## Methods

Cellular automata (CA) is implemented by creating a 2-dimensional cell grid of dimension 500×500. Each cell in the CA framework represents one of the seven possible cell states as defined in Figure 3A, *i.e.* at any time, $t$, any cell can assume one of the following states: susceptible (S), latent (L), quarantined (Q), infected (I), isolated (J), recovered (R) and deceased (D). Figure 3A presents the flow chart of all the possible interactions between the states, along with the probability of transitions. A detailed discussion regarding the relation between different states, their temporal transition and spatial distribution is provided in the supplementary information. While Figure 3B presents a representative snapshot of cellular states over a region, Figure 3C illustrates the smallest local neighbor, *a.k.a.* Moore's neighbor, which consists of nine neighbor cells. Details of the CA algorithm employed in the present work is provided in the supplementary document. We note that our predictions have an extremely low mean absolute error (MAE) of 0.0624.

**FIGURES**

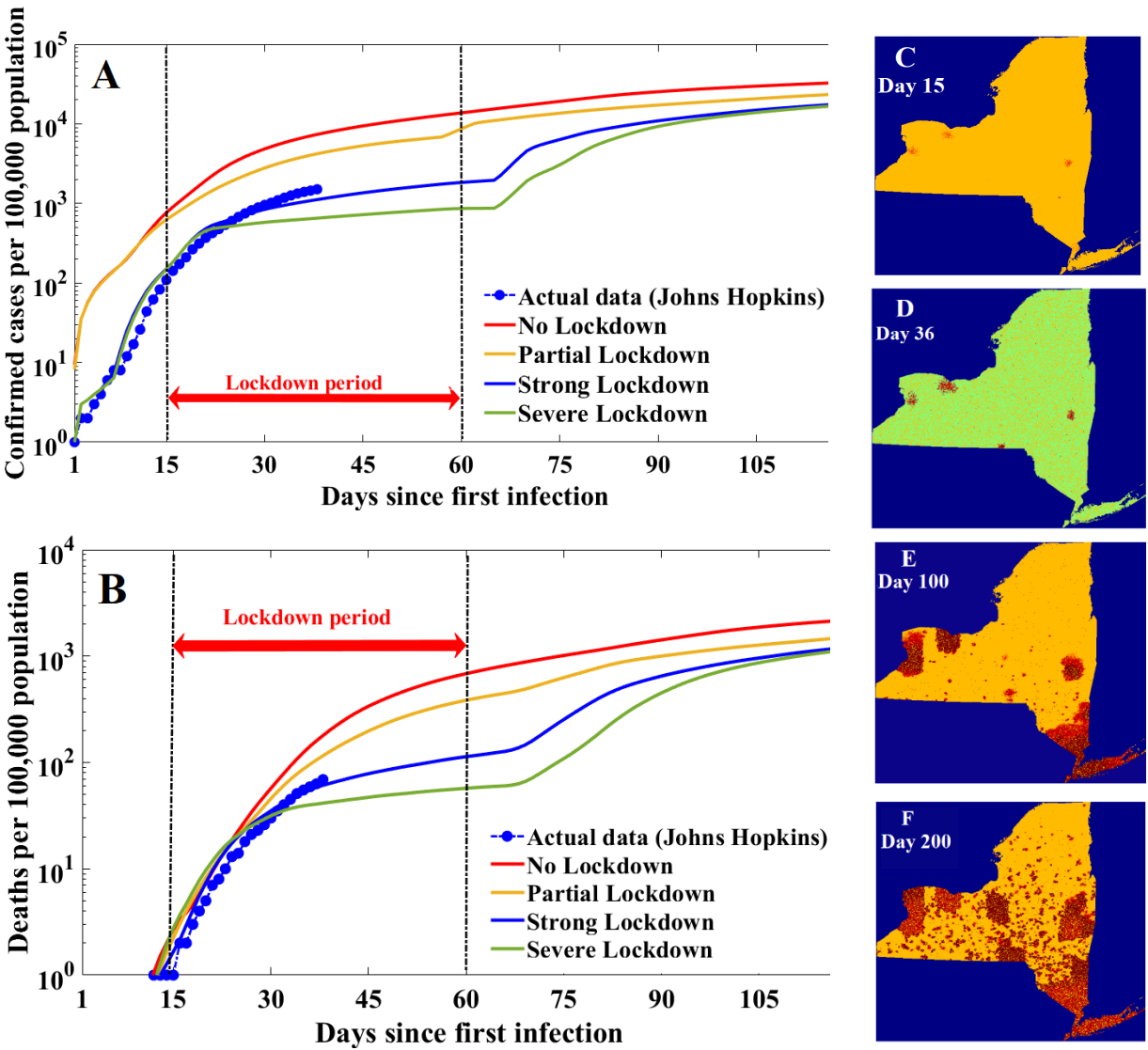

**Figure 1: Imposed lockdown and its effect on the outbreak in New York (NY). (A)** Cumulative number of confirmed cases per 100,000 population is presented as a function of time (in days) for four different scenarios *viz.,* no lockdown, partial lockdown (30%), strong lockdown (70%) and severe lockdown (95%) as predicted from the CA-SLQIJR model. Real time series data concurs with the strong lockdown scenario (as of 14th April 2020), since first infection out of 100,000 population was confirmed in NY state. **(B)** Cumulative number of deaths per 100,000 population is predicted as a function of time (in days) for the same four different scenarios as Figure (A). Real

time data of reported fatalities in NY state agrees with the predictions with the scenario where 70% of the population abides by the lockdown guidelines. **(C-F)** The hot spots for the disease spread (data scaled per 100,000 people) reproduced for four different days since the first reported infection illustrates that **(C)** few hot spots of disease outbreak existed at day 15 when the lockdown is first imposed. **(D)** An increase in the disease transmission is noted on day 36 (during the lockdown) reflecting the current situation in NY. The green color spread throughout the state denotes the population under self-quarantine. **(E)** Day 100 highlights the occurrences of new hotspots if the lockdown is relaxed after 45 days (*i.e,* 30$^{th}$ April, 2020). **(F)** An abrupt increase in the outbreak with another new wave of the epidemic is predicted after about 200 days.

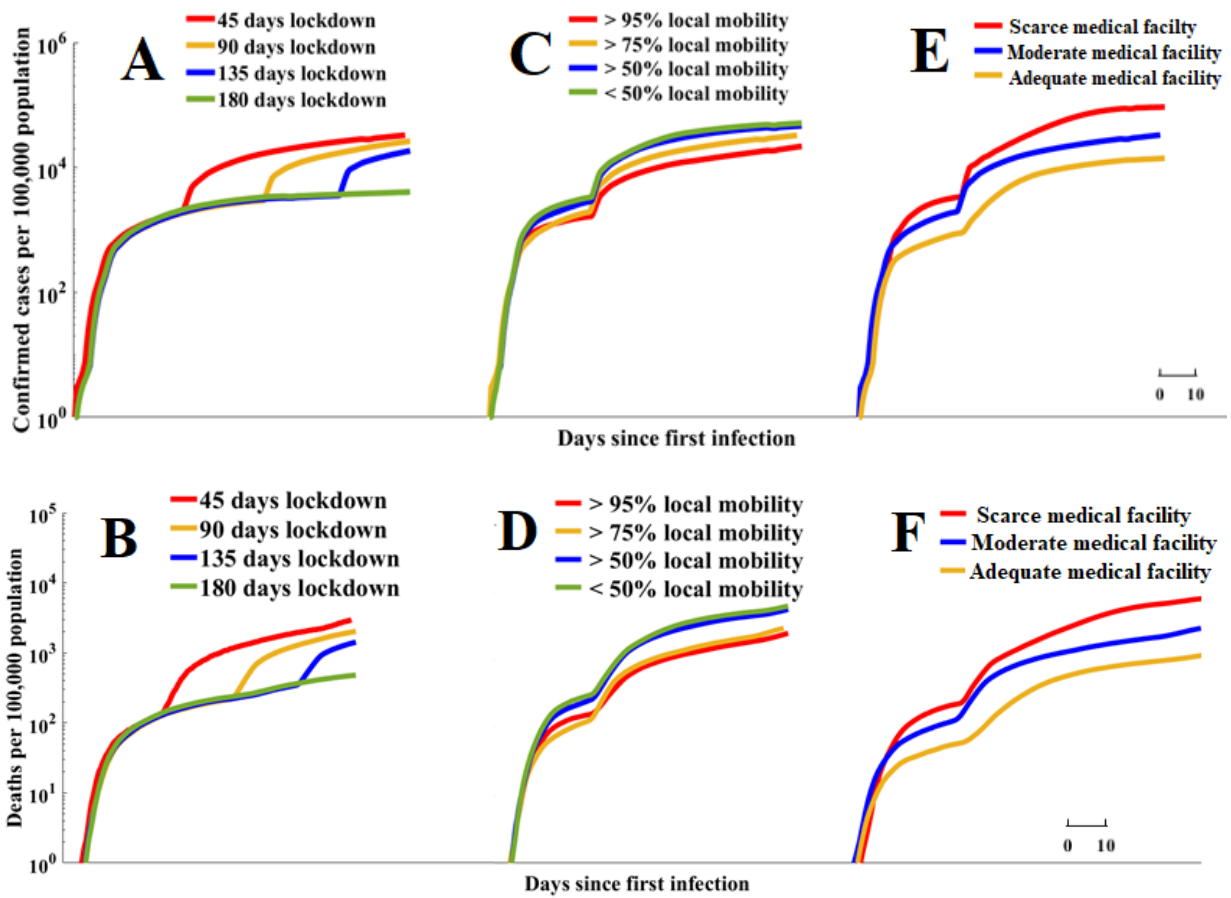

**Figure 2: Impact of key influential factors on the dynamics of disease transmission. (A) – (B)** The impact of extended lockdown on the total count of infections and fatalities are presented as a function of time (in days). An increase in lockdown period can reduce the total number of fatalities by an order of magnitude. **(C) – (D)** The impact of population mobility (local movement within a geography as opposed to long distance travel) on the cumulative number of infected cases and fatalities are reported as a function of time (number of days). Lockdown guidelines together with restricted movements within small communities can be beneficial to suppress the disease spread to other regions within NY. **(E) – (F)** The impact of available medical facilities on the cumulative number of infections and fatalities are interrogated for different scenarios: (1) adequate facility *i.e.,* > 95% of cases are identified and tested, (2) moderate facility *i.e.,* 75% of cases are tested and (3) scarce facility *i.e.,* only 25% cases are being tested. Adequacy of medical facilities can enable

rapid identification and isolation of infected cases and limit the damage to human life and economy.

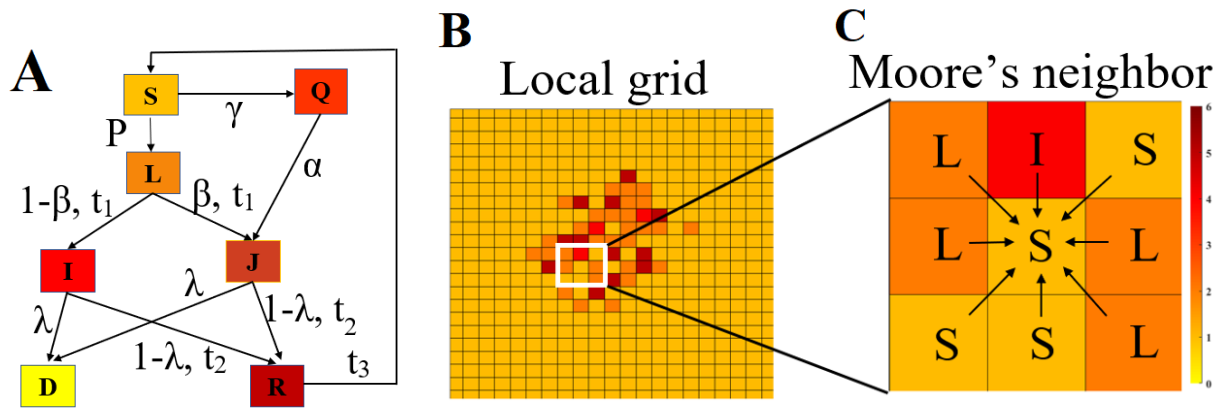

**Figure 3: Pictorial representation of the computational method. (A)** A flow chart describing the transition between different states from Susceptible (S) – Latent (L) – Quarantine (Q) – Infectious (I) – Isolated (J) – Recovered (R) – Death (D). Initially the entire population of NY is composed of susceptible people. One infected person (I) is injected in New York City area where first outbreak started. **(B)** Cellular automata grids are shown with the smallest possible Moore's neighbor **(C)** that contains nine neighbor cells. The cell at the center represents a susceptible person that is vulnerable to exposure from the eight Moore neighbors.

[20] The New York Times. Lockdowns in France and U.K. Expected to Last into Next Month. https://www.nytimes.com/2020/04/13/world/coronavirus-news-world-international-global.html (2020)

[21] Matthew, I. Which State Lockdowns in the U.S. Have Been Extended? https://www.newsweek.com/which-state-lockdowns-us-have-been-extended-1497162 (2020)

[22] Lau, H. et al. The Positive Impact of Lockdown in Wuhan on Containing the COVID-19 Outbreak in China. ***J Travel Med.***, Taaa037 (2020)

# SUPPLEMENTARY INFORMATION

# Spatiotemporal dynamics in demography-sensitive disease transmission: COVID-19 spread in NY as a case study

Joydeep Munshi, Indranil Roy and Ganesh Balasubramanian[Γ]

*Department of Mechanical Engineering & Mechanics, Lehigh University, Bethlehem, PA 18015*

**Computational Methods**

We employ coupled CA-SLQIJR model that predicts the impact of population dynamics such as density, mobility, and access to public health, on the ongoing pandemic situation due to the COVID-19 disease outbreak. In order to reproduce the effect of population density, we discretize the simulation domain into 500×500 cell grids. We sample the results for all the scenarios to reduce the overall noise due to the local random walks of the CA cells mimicking the population mobility within a geographic region. The code is scalable with additional demographic inputs and can be implemented for any city, state or country. We consider the CA cells as individual elements of a big simulation box with fixed dimensions where each cell can change its state according to state transition rules defined by the SLQIJR model, *i.e.* the value of an element and its position will change over time based on a certain defined ruleset. This 2-dimensional CA is implemented using differential equations to account for the temporal variation and the mobility of the population within different locations as spatial variations. The cellular automata model can be summarized as,

$$C = (L, S, N, f) \qquad (1)$$

Here, $L$ is the mesh dimension or spatial distribution, $S$ is the possible state of cells, $N$ denotes the neighbor of each cell and $f$ is the state transition function. In case of the COVID-19 outbreak, the possible states of an individual are S (susceptible to the disease), L (latent, i.e. infected or exposed but not yet symptomatic), Q (quarantine), I (infective), J (isolated), R (recovered) or D (dead) (SLQIJR). This classification follows from an earlier model popularly known as SEQIJR (Susceptible-Exposed-Quarantine-Infective-Isolation-Recovered). For all epidemic simulations we assume (1) no vaccination is invented during the entire simulation time and (2) people in the recovered (R) state can be re-infected for a second time once their immunity is gradually diminished with time. As we are not certain about the time period for loss of immunity (Table S1) we consider it to be on average ~3-6 months based on recently published reports. The governing differential equations for temporal transition of different states are

$$\frac{dS}{dt} = \tau_3 R - \delta\, SI - \gamma\, S \qquad (2)$$

$$\frac{dL}{dt} = \delta\, SI - \tau_1 L \tag{3}$$

$$\frac{dQ}{dt} = \gamma\, S - \tau_4\, Q - \alpha\, Q \tag{4}$$

$$\frac{dI}{dt} = (1-\beta)L - \tau_2\, I - \lambda\, I \tag{5}$$

$$\frac{dJ}{dt} = \alpha\, Q + \beta\, L - \tau_2 J - \lambda J \tag{6}$$

$$\frac{dR}{dt} = (1-\lambda)(I+J) - \tau_3 R \tag{7}$$

Here, $\delta\, I$ is the fraction of infected people influencing any susceptible (S) population and is determined based on a probability (P) as discussed later in the equations (8) and (9), $\gamma$ is the fraction of people in Quarantine state (Q), $\tau_1$ is the probability of a latent (L) person will be infective ($\sim \frac{1}{t_1}$), $\tau_2$ is the probability of an infective (I) or isolated (J) person will recover ($\sim \frac{1}{t_3}$), $\tau_3$ is the probability ($\sim \frac{1}{t_3}$) that a recovered person will lose immunity against the virus, $\gamma$ is the fraction of population following the imposed social restrictions such as lockdown (Q), $\beta$ is the fraction of latent population moving into isolation (J), $\alpha$ is the fraction of self-quarantined population becoming infected or symptomatic and subsequently being isolated (J) and finally $\lambda$ is the mortality fraction among the total infected population.

While the differential equations in SLQIJR model is used to determine the state transition function of each cell over time, spatial interactions are assumed by considering Moore's neighbors where eight surrounding cells are assumed as the neighbors of a cell. The state transition function (*f*) is determined by the transitional function (differential equations (2)-(7)) and probabilities of change of state as defined in Tables S1-S4. Figure 3A explains these transitions and the corresponding probabilities. In our model these probabilities are determined using multiple influential factors, making it sensitive to the population density and mobility, socio-economic background, age group, pre-existing health condition, medical facilities and employment type. Tables S1-S4 present all the parameters that are considered as part of our model to investigate disease spread as a function of the different influencing factors. The model is initialized with all susceptible candidate cells and one infected cell implanted in the New York city (NYC) where it is assumed to initiate the transmission due to substantially high population density and mobility. We employ the following algorithm in order for the model to propagate in time and space.

1. Initially all the cells in the simulation domain are S as everyone in a state can be part of the susceptible group if they are exposed to an infected person.
2. If a quarantine state (lockdown or stay at home) is imposed, cells with S states shall change their status to Q with a probability γ. Additionally random movements of rest of the susceptible crowd will slow down as part of the imposed lockdown with the same probability.
3. Initially a particular number of infectious (I) person are assigned to some of the cells. This data is retrieved from real time series information (Johns Hopkins University) of NY[1]. We assume our start day to be same as the date when there was one infected person reported in NY for every 100,000 population.

4. For a cell (i, j) probability P for the disease transmission is calculated when a susceptible cell (S) is in the neighborhood of an infected cell (I), as can be observed in Figure 3C. The probability depends on a person's own immunity (known as resistivity) and the state of neighboring cells (how many of the eight neighbors are infected). If the value of P is higher than a threshold, the state of cell (i, j) changes from S to L (latent).
We define P as

$$P_{(i,j)} = \frac{1}{4} \sum_{k \neq i \text{ and } l \neq j}^{8 \text{ neighbours}} \left[ \sqrt{fc_{(i,j)}, c_{(k,l)} (1 - R_{c(i,j)})} \right]$$

$$+ \frac{1}{4} \sum_{k = i \text{ and } l = j}^{8 \text{ neighbours}} \left[ \sqrt{fc_{(i,j)}, c_{(k,l)} (1 - R_{c(i,j)})} \right] \quad (8)$$

Where, $f$ is a random distribution function.

$R_{c(i,j)}$ is a person's resistance to infection, which is defined as

$$Rc_{(i,j)} = [af_1 \ af_2 \ af_3] \begin{bmatrix} aw_3 \\ aw_3 \\ aw_3 \end{bmatrix} [hf_1 \ hf_2 \ hf_3] \begin{bmatrix} hw_3 \\ hw_3 \\ hw_3 \end{bmatrix} [ef_1 \ ef_2 \ ef_3] \begin{bmatrix} ew_3 \\ ew_3 \\ ew_3 \end{bmatrix} T_c \quad (9)$$

$R_c$ is influenced by the age group distribution in the region and their susceptibility to the disease (a), fraction of population having pre-existing conditions which makes them susceptible to the COVID-19 (h) and employment type that can enable a faction of people to be vulnerable to the infection (e). These factors are accounted for by the dot product of fractions ($f$) and their respective weights ($w$). $T_c$ is the transferability of an infectious person present in the surroundings that we consider as a uniform distribution[2].

5. If the state of a cell, $C_{(i,j)}$ is L, a timer $T^s_{(i,j)}$ begins for the cell until it reaches higher than a defined value (*a.k.a.* latency period) denoted by $t_1$ days. At this point the state $C_{(i,j)}$ moves from L to I (or J), depending on the isolation probability (β).
6. Similarly, for cell states with I and J, it can either recover (R) or die (D) after a certain time, $t_2$ (recovery period) depending on the probability called morbidity (λ).
7. Finally, a recovered state with R transits to S as it again becomes susceptible once its immunization is lost (loss of immunity) after a certain time $t_2$ (immunity period).
8. Even from Q, a percentage α will shift to isolation (J) for those who get the disease. This count corresponds to the fraction of people who are infected when in self-quarantine. This assumption is based on a recent report suggesting that uncounted fatalities are reported in NY who died at home without getting proper medical attention.[3]

9. Finally, at each time step, only cells containing S, I and R states will move randomly based on a mobility parameter defined by the automata algorithm which is a function of local population density and effective lockdown imposed by the government.

**Tables**

**Table S1: Time periods for the different stages of the spread.** Any contagious disease such as COVID-19 has several stages such as latency period, infective period and recovery period. While we consider latency period and immunization period from known literature loss of immunity is not a well-known parameter.

| Characteristics time | Mean | Reference |
|---|---|---|
| Incubation or latency period ($t_1$) | 5.1 days | 4 |
| Immunization/recovery time ($t_2$) | 20.2 days | 5 |
| Lost immunity time ($t_3$) | ~ 90-180 days | |
| Lockdown period ($t_4$) | ~ 45-180 days | |

**Table S2: Age group as an influential factor affecting COVID-19 outbreak.** Distribution of age groups with resistance to infection are considered to study its effect of on the model prediction. Older and younger age groups are reported to be most vulnerable to the infection. Our model considers the statistical data of age group distribution of NY.

| Age Group (years) | Percentage of population[6] | Resistivity Percentage |
|---|---|---|
| 1-10 | 11 | 20 |
| 11-54 | 59.4 | 60 |
| 55 and higher | 29.6 | 20 |

**Table S3: Pre-existing health condition as an influential factor.** People with respiratory disease (*e.g.* asthma) are reported to be most vulnerable followed by those with cardiovascular diseases. We model the probability of fatalities and infections based on the statistics of pre-existing health conditions of NY.

| Condition | Percentage of population | Resistivity Percentage |
|---|---|---|
| Respiratory | 8 | 5 |
| Cardiovascular | 48 | 15 |
| None | 44 | 80 |

**Table S4: Employment[7] status of the population as an influential factor**. Fraction of people of a region who are responsible for emergency services are considered as an influencing factor in our model *e.g.* health care and emergency service providers are the most vulnerable.

| Type | Percentage of population | Resistivity Percentage |
|---|---|---|
| Health Care | 13.5 | 10 |
| Emergency Service | 15 | 20 |
| Other | 71.5 | 70 |

**Figures**

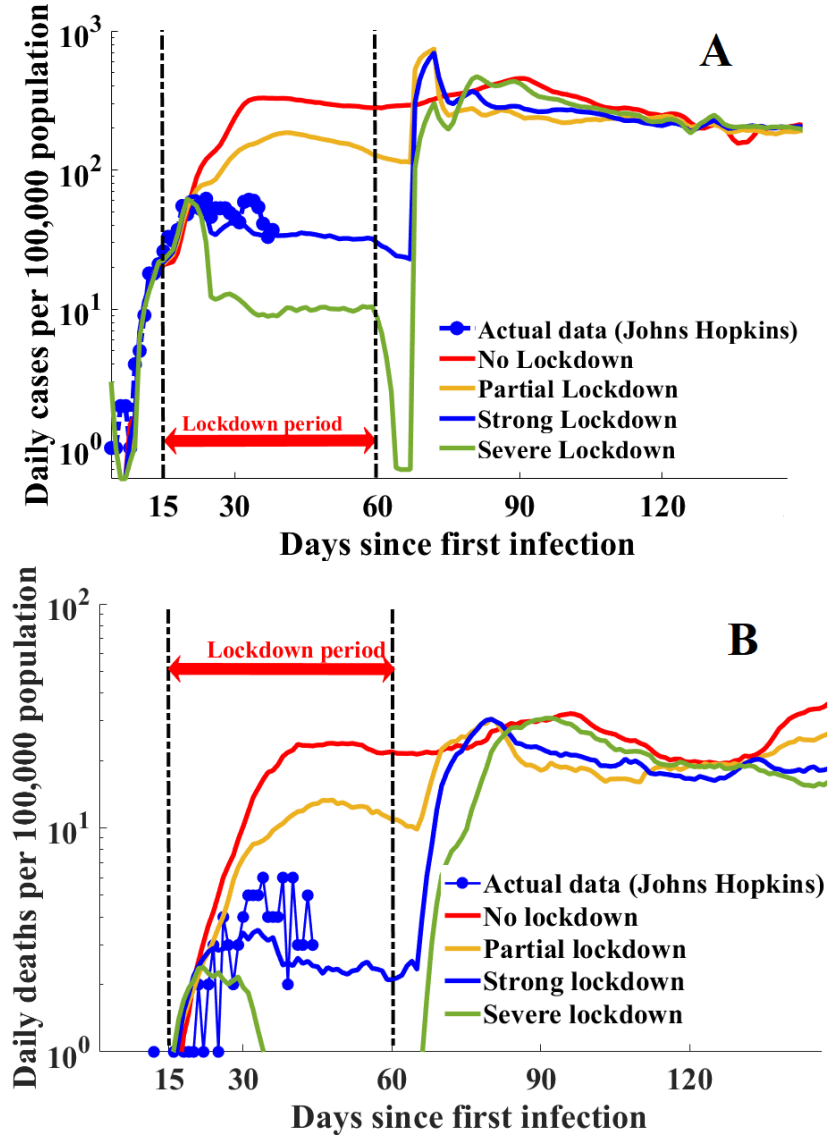

**Figure S1: Daily cases for four different scenarios (Log-scale). (A)** Daily predicted infection cases are presented for no lockdown, partial lockdown, strong lockdown and severe lockdown scenarios as a function of days since the first infection was detected. **(B)** Daily predicted death count is presented as a function of days since the first report. Both (A) and (B) reproduce good agreement of the strong lockdown scenario with the actual data available and published by JHU.

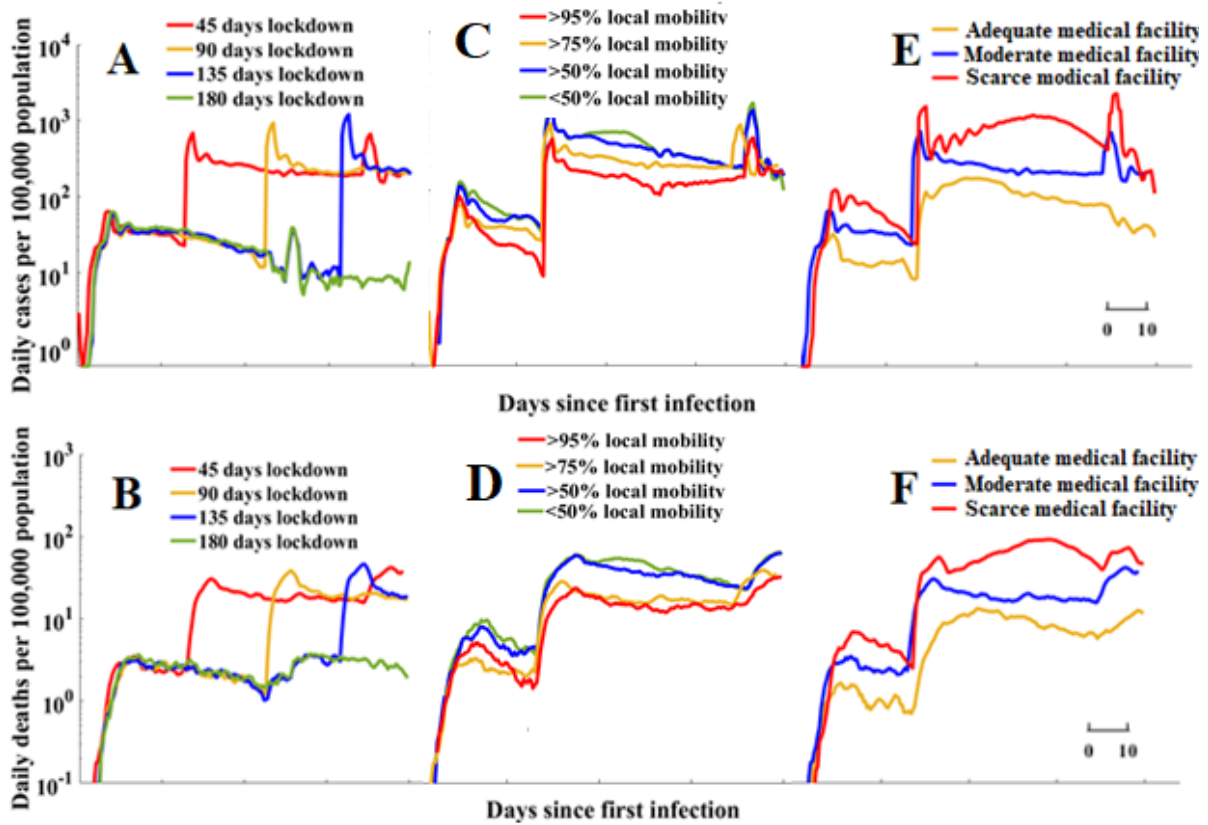

**Figure S2: Daily predictions for different influential factors (Log-scale). (A)-(B)** The impact of different lockdown periods is presented. Extended lockdown up to 180 days is found beneficial to reduce the total cases and fatalities. **(C)-(D)** Local movement of a fraction of the population impacts the total number of infections and death. < 50% mobility is observed to suppress the overall infections and deaths. **(E)-(F)** Daily predictions of infected cases and deaths are presented as a function available medical facility as in *adequate* (>95%), *moderate* (75%) and *scarce* (25%) scenarios. With increase in access to a medical facility, the total infections and deaths are noted to rapidly reduce potentially enhancing the overall confidence within the population.

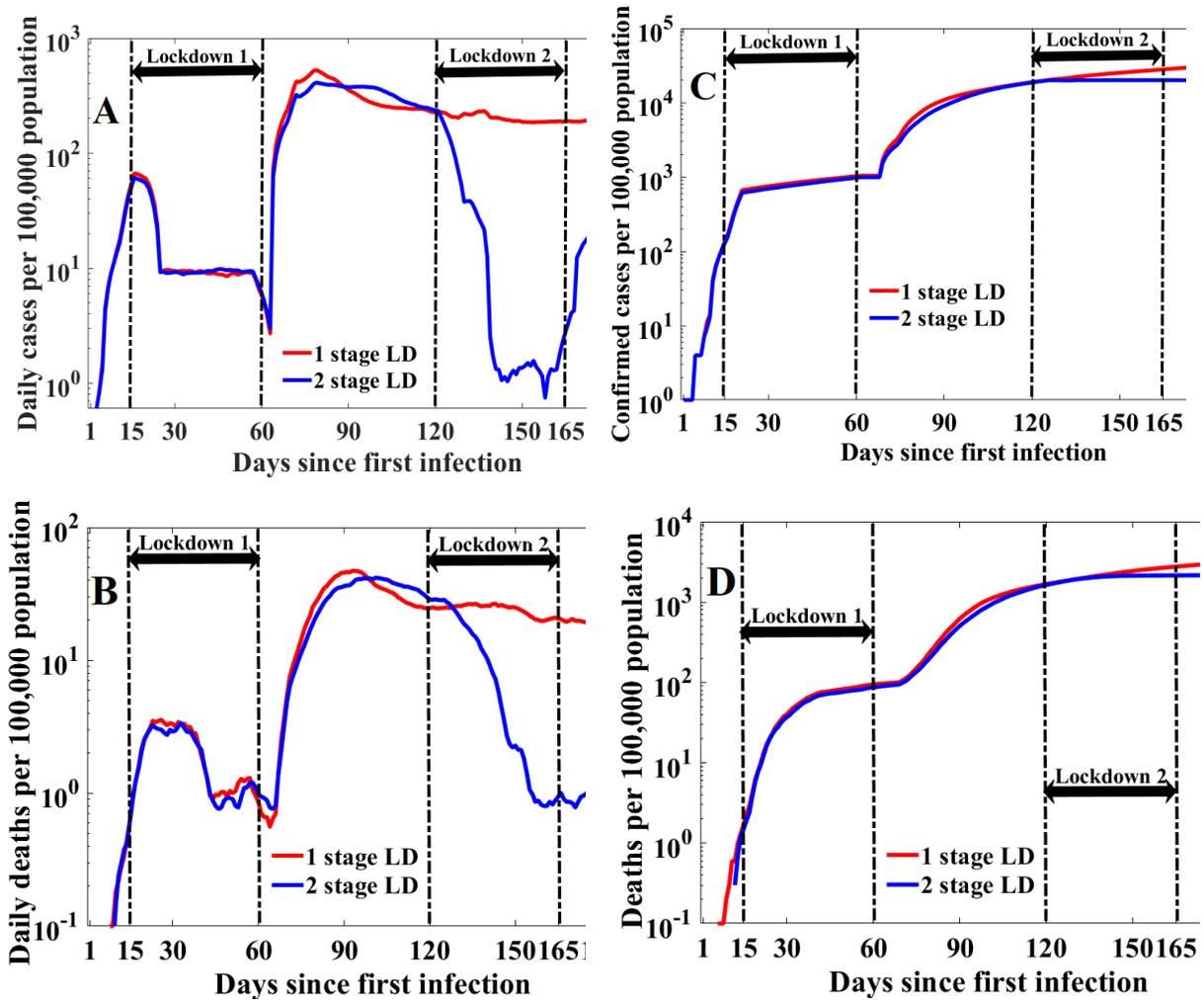

**Figure S3: Comparison of 1 stage and 2 stage lockdowns (Log-scale). (A)-(B)** Daily confirmed cases and deaths for 1 stage lockdown (15$^{th}$-60$^{th}$ day) and 2 stage lockdowns (15$^{th}$-60$^{th}$ and 120$^{th}$-165$^{th}$ day) reveal 2 stage lockdowns can be beneficial to mitigate the outbreak in a great extent. Here we consider complete lockdown (>99% of population observing self-quarantine). **(C)-(D)** Cumulative confirmed cases of infection and deaths are predicted as a function of days since the first reported infection. In the case of 2-stage lockdown, cumulative cases and deaths are observed to saturate significantly.